\documentclass[12pt]{article}
\usepackage{psfig}
\begin{document}
\begin{center}
%\begin{frontmatter}
{\Large {\bf Deuteron photodisintegration within the Quark-Gluon
Strings Model and QCD motivated nonlinear Regge
trajectories}}\footnote{Supported by DFG and RFFI.}
\end{center}
\vspace{2cm}
\begin{center}
V.Yu.~Grishina$^a$, L.A.~Kondratyuk$^b$, W.~Cassing $^c$,
A.B.~Kaidalov$^b$, E.~ De~Sanctis$^d$ and P.~Rossi $^d$ \\
\vspace{1cm}
$^a${\it Institute for Nuclear Research, 60th October
Anniversary
  Prospect 7A, 117312 Moscow, Russia}\\
$^b${\it Institute for Theoretical and Experimental Physics, B.\
  Cheremushkinskaya 25, 117259 Moscow, Russia}\\
$^c${\it Institute for Theoretical Physics,  University of
Giessen, Heinrich-Buff-Ring 16, D-35392 Giessen, Germany}\\
$^d${\it Frascati National Laboratories, INFN, CP 13, via E.
Fermi, 40; I-00044, Frascati, Italy}\\

\end{center}
\vspace{1cm}

\begin{abstract}
We investigate deuteron two-body photodisintegration within the
framework of the Quark-Gluon Strings Model with nonlinear baryon
Regge trajectories. Special attention is paid to the use of QCD
motivated Regge trajectories of the logarithmic and square-root
form which have been suggested recently by Brisudov\'{a},
Burakovsky and Goldman. We find that  the recent experimental data
from TJNAF in the few GeV region  can reasonably be described by
the model. Angular distributions at different $\gamma$-energies
are presented and the effect of a forward-backward asymmetry is
discussed. Predictions for the energy dependence of $d\sigma/dt$
at higher energies and different $\Theta_{cm}$ are presented, too.
\vspace{1cm}

%\begin{keyword}

\end{abstract}

\newpage
\section{Introduction}
\label{sec:qgsm}

Recent experiments on high energy two-body photodisintegration of
the deuteron \cite{Holt}-\cite{Belz} have brought up interesting
results: while the $89^{\circ}$ and $69^{\circ}$ data are
consistent with the constituent-quark counting-rule behavior
\cite{Matveev} (i.e. at fixed c.m. angle the differential cross
section $d\sigma/dt_{\gamma d \rightarrow pn}$ scales as $ \sim
s^{-11}$), the $36^{\circ}$ and $52^{\circ}$ data do not show a
scaling behavior at all up to $4.0$~GeV photon energy. Thus
perturbative QCD (PQCD) cannot be applied at these energies at
forward angles and nonperturbative approaches have to be used
instead.

\begin{figure}[htb]
\begin{center}
    \leavevmode
    \psfig{file=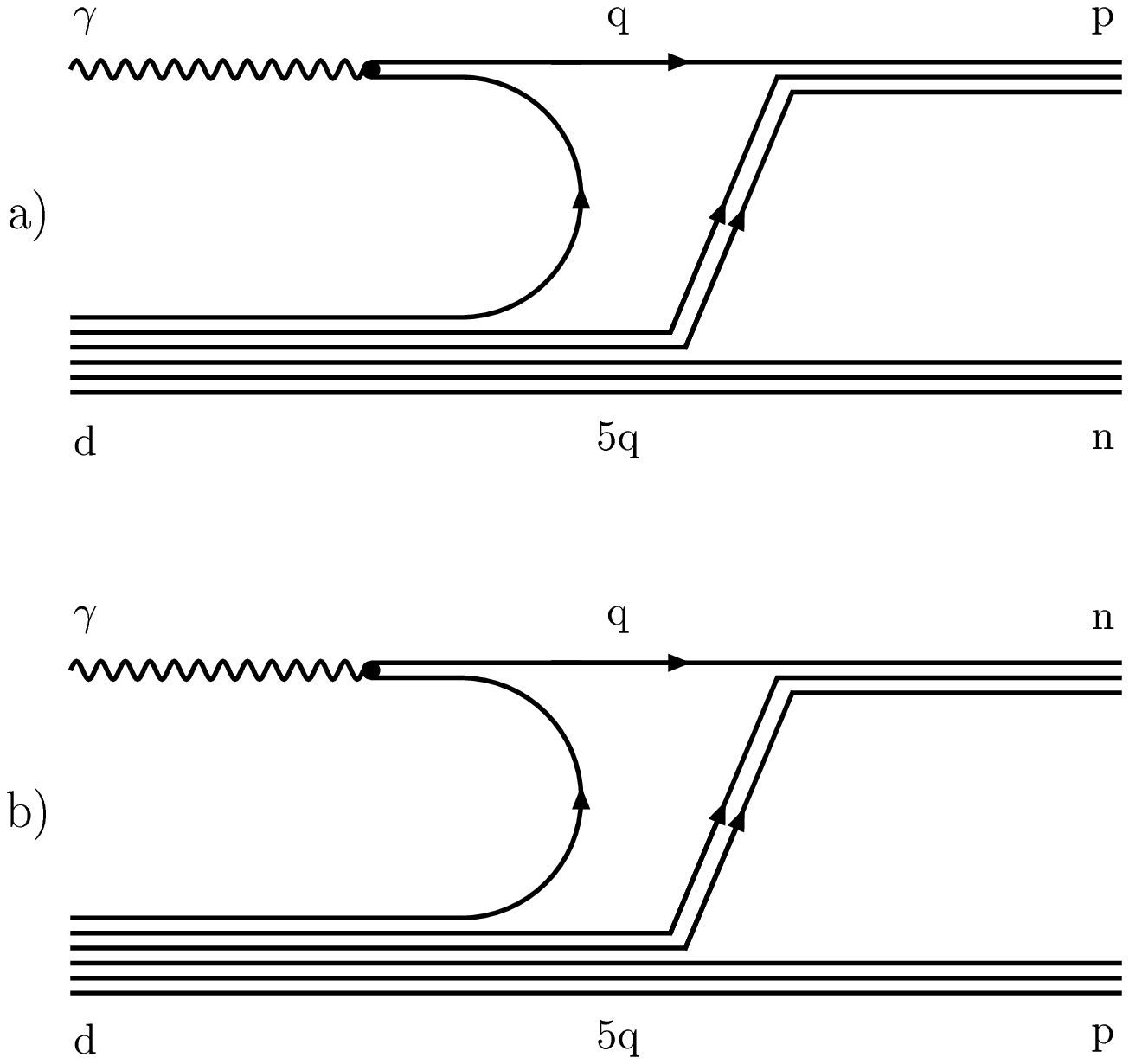,width=14.cm}
    \caption{Diagrams describing three valence quark exchanges in $t$- (a) and
      $u$-channels (b).}  \label{fig:qgsm}
  \end{center}
\end{figure}

Some time ago a nonperturbative approach based on the Quark-Gluon
Strings Model (QGSM) has been applied to the analysis of the
angular and energy dependence of the differential cross section
for the $\gamma d\to pn$ reaction in the few GeV energy region
\cite{Desanctis}. In the QGSM -- proposed in Ref. \cite{Kaidalov}
for the description of binary hadronic reactions -- the amplitude
of the reaction $\gamma d \to pn$ is described by the exchange of
three valence quarks in the $t$-channel with any number of gluon
exchanges between them. This process is visualized in Fig.
\ref{fig:qgsm}, where a) and b) describe the exchange of three
valence quarks in the $t$- and $u$-channels, respectively.

We recall that the QGSM is based on two ingredients: i) a
topological expansion in QCD and ii) the space-time picture of the
interactions between hadrons, that takes into account the
confinement of quarks.  The $1/N$ expansion in QCD (where $N$ is
the number of colors $N_c$ or flavors $N_f$) was proposed by 't
Hooft \cite{Hooft}; the behavior of different quark-gluon graphs
according to their topology, furthermore, was analyzed by
Veneziano \cite{Veneziano} with the result that in the large $N$
limit the planar quark-gluon graphs become dominant. This approach
based on the $1/N_f$ expansion \cite{Veneziano} with $N_c \sim
N_f$  was used by Kaidalov \cite{Kaidalov,KaidalovSurvey} in the
formulation of the QGSM. Again for sufficiently large $N_f$ the
simplest planar quark-gluon graphs give the dominant contribution
to the amplitudes of binary hadronic reactions. Moreover, it can
be shown that in the space-time representation the dynamics
described by planar graphs corresponds to the formation and
break-up of a quark-gluon string (or color tube) in various
intermediate states (see e.g. Refs. \cite{Casher, Artru, Casher2,
Andersson, Gurvich}). Here the quark-gluon string can be
identified with a corresponding Regge trajectory\footnote{In case
of Fig.\ref{fig:qgsm} we have in the intermediate state a string
with a quark and a diquark at the ends which corresponds to a
nucleon Regge trajectory}. In this sense the QGSM can be
considered as a microscopic model of Regge phenomenology and be
used for the calculation of different parameters, that have been
considered before only on a phenomenological level.

As shown in Refs. \cite{Kaidalov,KaidalovSurvey} the QGSM
describes rather well the experimental data on exclusive and
inclusive hadronic reactions at high energy.  Moreover, due to the
duality property of scattering amplitudes this approach can also
be applied at intermediate energies for reactions without explicit
resonances in the direct channel. In fact, this model successfully
describes the reactions $pp \to d \pi^+$, $\bar{p}d\to MN$ and
$\gamma d \to pn $ at intermediate energies, too, where the
diagrams with three valence quark exchanges in the $t$-channel
were found to be dominant (cf. Refs.
\cite{Desanctis,Kaidalov4,Guaraldo}). However, in all those cases
the explicit spin structure of the corresponding amplitudes was
not taken into account. On the other hand, spin effects and the
transversal polarization of the photon lead to a nontrivial
angular dependence of the residue of the amplitude for the
reaction $\gamma d \to pn $ as discussed in Ref. \cite{Desanctis}.
In this paper we will use an extended approach to the spin effects
in the QGSM as developed in Ref. \cite{Chekin} with respect to the
description of the electromagnetic nucleon form factors $F_1$ and
$F_2$.

Another important extension of the QGSM in our study will be the
use of nonlinear baryon Regge trajectories. There are
phenomenological evidences (see \cite{Lyubimov,UA8,Inopin}) as well as
theoretical arguments (see e.g. \cite{Burak} and references
therein) that hadronic Regge trajectories should be nonlinear.
This nonlinearity is not important for small momentum transfer
(squared) $t$, however, in the region of $t$, which has already
been reached in the experiment \cite{Holt},  effects of
nonlinearity become very essential. We will employ three different
forms of nonlinear nucleon Regge trajectories: i) a
phenomenological one, which becomes linear at large $t$, and two
QCD motivated trajectories of ii) logarithmic and iii) square-root
type as suggested recently by Brisudov\'{a}, Burakovsky and
Goldman \cite{Burak}. By extrapolating our amplitudes to large
angles we can figure out to what momentum transfer $t$ the 'soft
QGSM tails' are still important before the PQCD regime becomes
dominant.

In Section 2 we outline the QGSM and its relation to Regge theory
and introduce hadron-quark and quark-hadron transition amplitudes
that are described by planar graphs. In Section 3 we construct the
$\gamma d \to p n $ amplitude taking into account explicit spin
variables. In Section 4 we present the results of our calculations
in comparison to the data \cite{Holt} while Section 5 concludes
our present study.

\section{The Quark-Gluon Strings Model}
\label{model}

In order to introduce the basic features of the QGSM we consider
the binary reactions $\pi ^{0}\pi ^{0}\rightarrow \pi^{+} \pi ^{-}$,
$ N \overline{N}\rightarrow\pi ^{+} \pi^{-}$ and  $p
\overline{p}\rightarrow N\overline{N}$, which at large values of
the invariant energy (squared) $s$ and finite values of the
4-momentum transfer (squared) $t$ can be described by  planar
diagrams with $t$-channel va\-lence-quark exchanges as shown by
the diagrams a)-c) in Fig. {\ref{fig:piNNbar}}. Here the single
and double solid lines correspond to valence quarks and diquarks,
respectively,  while soft gluon exchanges between these lines are
not shown. According to the topological $1/N_f$ expansion
\cite{Veneziano,KaidalovSurvey,Kaidalov1}, these planar diagrams
are
expected to give the dominant contributions to the corresponding
amplitudes in the limit $N_f\gg 1$ and $N_c/N_f\sim 1$. In the
case of pion and nucleon interactions -- as considered here -- the
exchanges of light $u$, $d$, and $s$ quarks are mainly important
and the parameter of the expansion is not small, i.e. $1/N_f=1/3$.
However, in case of amplitudes for exclusive reactions with
specific quantum numbers in the $t$-channel, the actual expansion
parameter is $1/N_f^2 \sim 1/9$ such that the expansion is
expected to work.

\subsection{Transition probabilities}
\begin{figure}
 \begin{center}
    \leavevmode
    \psfig{file=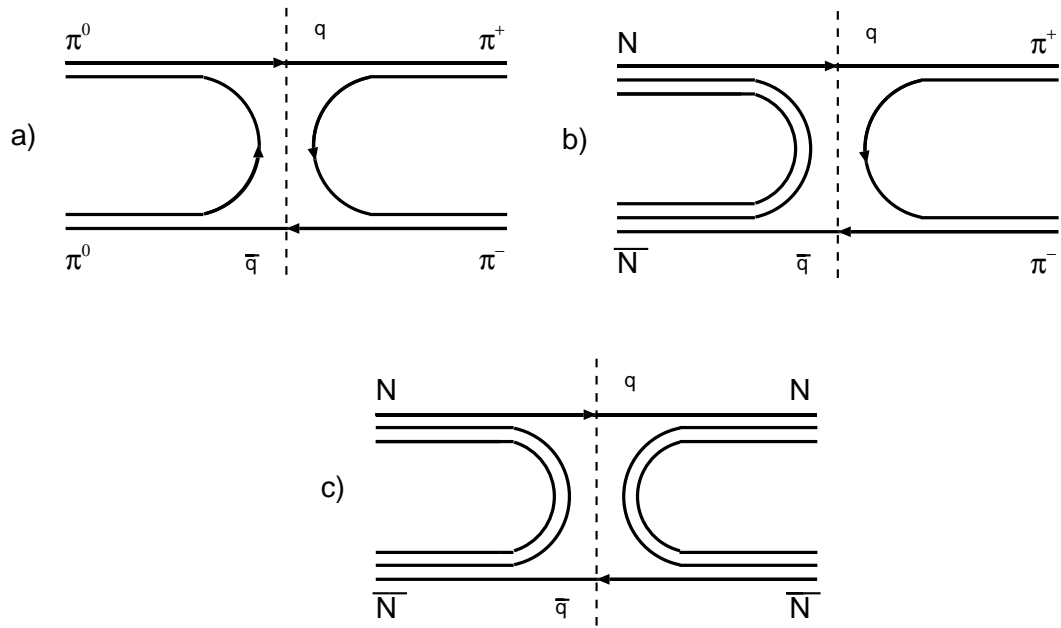,width=14.cm}
    \caption{Planar diagrams describing the binary reactions
a) ${\pi}^{0} {\pi}^{0} \rightarrow {\pi}^{+} {\pi}^{-}$, b) $N
\bar {N} \rightarrow {\pi}^{+} {\pi}^{-} $, c) $p \bar {p}
\rightarrow  N \bar {N}$.} \label{fig:piNNbar}
  \end{center}
\end{figure}

Each planar diagram of the topological expansion has a simple
interpretation within the framework of the space-time pattern
formulated in terms of a color tube (or color string)
\cite{KaidalovSurvey,Kaidalov3}. As an example we consider the
space-time picture
of the binary reaction $\pi ^{0}\pi^{0}\rightarrow \pi^{+} {\pi }^{-}$
( cf. Fig.{\ref{fig:piNNbar}} a) ). At high center-of-mass (c.m.)
energy $\sqrt{s}$, this reaction occurs due to a  specific quark
configuration in each pion, where  (in the c.m. system) one quark
(antiquark) takes almost the entire hadron momentum and plays the
role of a spectator, while the valence antiquark (quark) is rather
slow. The difference in the rapidities $\Delta y$ between the
quark $q$ and antiquark $\bar q$ in each pion is
\begin{equation}
\label{rapidity}\Delta y = y_q-y_{\bar q}\simeq \frac12\ln \left(
\frac s{s_0}\right),
\end{equation}
with the scale $s_0 \simeq 1$ GeV$^2$. Then the two 'slow' valence
partons $q$ and $\bar q$  from $\pi ^{0}$ and $\pi ^{0}$
annihilate, and the fast spectator quark and antiquark continue to
move in the previous directions and form a color string in the
intermediate state. After that, the string breaks due to the
production of a $q \bar q$-pair from the vacuum and formation of
the $\pi ^{+} \pi ^{-}$ system in the final state. We note, that
the same space-time pattern holds for the diagram of Fig.
{\ref{fig:piNNbar}} b) with the only difference, that the string
is formed after annihilation of a diquark-antidiquark pair from
the $N\overline{N} $-system in the initial state. Correspondingly,
the graph of Fig. {\ref{fig:piNNbar}} c) shows the formation of
the $q\overline{q}$ string due to annihilation of the valence
diquark-antidiquark pair in the initial state and the production
of another diquark-antidiquark pair due to the breaking of the
string.

The annihilation of the initial $q\overline{q}$ (or $(qq)
(\overline{qq})$) pair takes place, when the gap in  rapidity of
the valence $q$ and $\overline{q}$ (or  $(qq) (\overline{qq})$) is
small (both interacting partons are almost at rest in c.m.s.) and
the relative impact parameter $\vert {\bf b}_{\perp} -{\bf
b}_{0\perp} \vert$ is less than their interaction radius. It is
possible to prove that the
probability to find a valence quark with a rapidity $y_q$ at
impact parameter ${\bf b}_{\perp }$ inside a hadron can be
written  as \cite{Kaidalov,KaidalovSurvey,Kaidalov3}
\begin{equation}
\label{probability} w\left(y_q-y_0,{\bf b}_{\perp}-{\bf
b}_{0\perp}\right)= \frac c{4\pi R^2(s)}\exp \left[ -\beta
(y_q-y_0)-\frac{({\bf b}_{\perp}-{\bf
b}_{0\perp})^2}{4R^2(s)}\right] \ ,
\end{equation}
where $c$ is a normalization constant, $y_0$ is the average
rapidity, ${\bf b}_{0\perp}$ is the transverse coordinate of the
c.m. system in the impact-parameter representation. Furthermore,
it is possible to relate the parameter $\beta$ and the effective
interaction radius squared $R^2(s)$ in (\ref{probability}), that
specify the quark distribution inside a hadron, to the
phenomenological parameters of a Regge trajectory $\alpha _i (t)$
which gives the dominant contribution to the amplitude for the
considered planar graph.  In this case one gets
\begin{equation}
\label{r2} R^2(s)=R_0^2+\alpha ^{\prime }_i (y_q-y_0)\ ,\quad
\beta =1-\alpha_i(0) \ ,
\end{equation}
where $\alpha_i ^{\prime }=\alpha _i^{\prime }(0)$ is the slope of
the dominant Regge trajectory.

Due to the creation of a string in the intermediate state the
amplitude of a binary reaction $ab\to cd$ has the $s$-channel
factorization property, i.e. the probability for the string to
produce different hadrons in the final state does not depend on
the type of the annihilated quarks and is only determined by the
flavours of the produced quarks. The same independence also holds
for the production of the color string in the intermediate state
from the initial hadron configuration: it depends only on the type
of the annihilated quarks. This $s$-channel factorization has been
formulated in Refs. \cite {Kaidalov}, \cite{KaidalovSurvey},
\cite{Kaidalov3} in terms of
transition probabilities as defined by Eq. (\ref{probability}).

\subsection{Transition amplitudes}
Following Ref. \cite{Chekin} we now generalize  this approach by
introducing the amplitudes $\widetilde{T}^{ab\to q\bar q} (s,{\bf
b}_{\perp})$ and $\widetilde{T}^{q\bar q\to cd}(s,{\bf b}_{\perp
})$, that describe the formation and the fission of an
intermediate string, respectively. The amplitude for the binary
reaction $ab \to cd$ described by the planar graph of Fig.1a) ( b)
or c)) can be written -- employing the $s$-channel factorization
property -- as a convolution of two amplitudes, i.e.
\begin{equation}\label{ImpRepr}
A^{ab\rightarrow cd}\left(
s,{\bf q}_{\perp }\right) = \frac i{8\pi ^2s}\int d^2{\bf k}_{\perp } \
T^{ab\rightarrow q\overline{q}} \left( s,{\bf k }_{\perp }\right)
T^{q\overline{q}\rightarrow cd}\left( s,{\bf  q }_{\perp }-{\bf
k}_{\perp }\right)
\end{equation}
in momentum representation, or as the product
\begin{equation}\label{factorization}
\widetilde{A}^{ab\rightarrow cd}(s,{\bf b}_{\perp })=\frac
i{2s}\ \widetilde{T}^{ab\to q\bar q}(s,{\bf b}_{\perp })\ \widetilde{T}
^{q\bar q\to cd}(s,{\bf b}_{\perp })
\end{equation}
in impact-parameter representation.

The solution for the quark-hadron transition amplitudes $T^{ q
\overline{q} \rightarrow \pi \overline{\pi } }\left( s,{\bf
k}_{\perp }\right)$ and $T^{q \overline{q}\rightarrow
N\overline{N}}\left( s,{\bf k}_{\perp }\right) $ at large
invariant energy $\sqrt{s}$ can be found using single Regge-pole
parameterizations  of the binary hadronic amplitudes $A^{\pi
^{0}\pi ^{0}\rightarrow \pi^{+} \pi^{-}}$,
$A^{N\overline{N}\rightarrow \pi \overline{\pi}}$ and
$A^{N\overline{N}\rightarrow N\overline{N}}$
\begin{equation}
 \label{MBDReggeAmplitude}
\begin{array}{c}
\displaystyle A^{\pi ^{0} \pi ^{0}\rightarrow \pi
^{+}\pi ^{-}}\left( s,t\right) =N_M\ \left( -\frac
s{m_0^2}\right) ^{\alpha _M\left( t\right) }\exp \left( R_{0M}^2t\right), \\
\displaystyle A^{ N
\overline{N}\rightarrow \pi \overline{\pi } }
\left( s,t\right) =N_B\ \left( -\frac
s{m_0^2}\right) ^{\alpha _B\left( t\right) }\exp \left( R_{0B}^2t\right), \\ \\
\displaystyle A^{N\overline{N}\rightarrow
N\overline{ N }}\left( s,t\right) =N_D\ \left( -\frac
s{m_0^2}\right) ^{\alpha _D\left( t\right) }\exp \left( R_{0D}^2t\right).
\end{array}
\end{equation}
Here $\alpha _M\left( t\right) $, $\alpha _B\left( t\right) $ and
$\alpha_D\left( t\right) $ are the dominant meson, baryon and
diquark-antidiquark trajectories while $N_M$, $N_M$ and $N_D$ are
normalization constants; $m_{0}^2=s_0$ and $R_{0 i}$ is the
interaction radius for the i-th trajectory. We have the following
intercepts and slopes for the dominant Regge trajectories
\begin{equation}
\label{ReggePoles}\alpha _M\left( 0\right) \simeq 0.5,\quad \alpha _B\left(
0\right) \simeq -0.5,\quad \alpha _D\left( 0\right) \simeq -1.5
\end{equation}
and
\begin{equation}\label{ReggePoles2}
\alpha _M^{\prime }\left( 0\right) \simeq \alpha _B^{\prime }\left( 0\right)
\simeq \alpha _D^{\prime }\left( 0\right) \simeq 1.0\ GeV^{-2}   .
\end{equation}
Using equations (\ref{factorization}) and (\ref{MBDReggeAmplitude}) we
can write  the amplitudes $\widetilde{T}^{q\overline{q}
\rightarrow \pi \overline{\pi}}\left( s,{\bf b}_{\perp }\right) $
and $\widetilde{T}^{q \overline{q}\rightarrow N\overline{N}}\left(
s,{\bf b} _{\perp }\right)$ as
\begin{equation} \label{PionNucleonFragmentation}
\begin{array}{c}
\displaystyle \widetilde{T}^{q\overline{q} \rightarrow \pi \overline{\pi
}}(s,  {\bf b}_{\perp})=N_M^{1/2}\frac 1{2\sqrt{\pi }R_M\left( s\right)
}\left( -\frac s{m_0^2}\right) ^{\left( \alpha _M\left( 0\right) +1\right)
/2}\exp \left( -  \frac{{\bf b}_{\perp }^2}{8R_M^2\left( s\right) }\right), \\
\\
\displaystyle \widetilde{T}^{q\overline{q}\rightarrow N \overline{N}}(s,{\bf
b}_{\perp})= N_D^{1/2}\frac 1{2\sqrt{\pi }R_D\left(s\right) }
\left( -\frac s{m_0^2}\right)^{\left( \alpha _D\left( 0\right)+1\right) /2}
\exp \left( -\frac{{\bf b}_{\perp }^2}{8R_D^2\left( s\right)}\right),
\end{array}
\end{equation}
where $R_M\left( s\right) $ and $R_D\left( s\right) $ are the
effective interaction radii given by
\begin{equation}\label{Radii}
\begin{array}{c}
\displaystyle R_M^2\left( s\right) =R_{0M}^2+\alpha _M^{\prime }\left( 0\right) \ln \left(
-\frac s{m_0^2}\right),  \\ \\
\displaystyle R_D^2\left( s\right) =R_{0D}^2+\alpha _D^{\prime }\left( 0\right) \ln \left(
-\frac s{m_0^2}\right).
\end{array}
\end{equation}
Now substituting the amplitudes (\ref{PionNucleonFragmentation})
into the factorization formula (\ref{factorization}) we get:
\begin{equation}\label{CrossAmplitude}
\begin{array}{c}
\displaystyle \widetilde{A}^{N\overline{N}\rightarrow \pi \overline{\pi }}
(s,{\bf b}_{\perp })= \\ \\
\displaystyle  \left( N_MN_D\right) ^{1/2}
\frac 1{4\pi R_D\left( s\right) R_M\left(  s\right)
}\left( -\frac s{m_0^2}\right) ^{\frac 12\left( \alpha _D\left(
0\right) + \alpha _M\left( 0\right) \right) }
\exp \left[ -{\bf b}_{\perp }^2 \left(
\frac 1{8R_M^2\left( s\right) }+\frac 1{8R_D^2\left( s\right) }\right) \right].
\end{array}
\end{equation}
For consistency of eqs. (\ref{CrossAmplitude}) and
(\ref{MBDReggeAmplitude}) we have to require the following
relations between the Regge parameters and normalization constants
(\cite{Kaidalov,KaidalovSurvey,Kaidalov1}):
\begin{equation}\label{OurPlanar} \begin{array}{c}
\displaystyle 2\frac 1{R_B^2\left( s\right) }=\frac 1{R_M^2\left( s\right)
}+\frac 1{R_D^2\left( s\right) } , \\ \\
\displaystyle 2\alpha \left( 0\right) _B=\alpha _D\left( 0\right) +\alpha _M\left(
0\right) ,
\end{array}
\end{equation}
\begin{equation}
\label{NormCnst}
\left( N_M N_D\right) ^{1/2}\frac 1{R_D\left( s\right)
R_M\left( s\right) }=N_B\frac 1{R_B^2\left( s\right) } .
\end{equation}
If only light $u$, $d$ quarks are involved we can assume that
(\cite{Kaidalov,KaidalovSurvey,Kaidalov1})
\begin{equation}\label{UDPlanar}
\begin{array}{c}
\displaystyle \alpha _M^{\prime }\left( 0\right) =\alpha _B^{\prime }\left(
0\right) =\alpha _D^{\prime }\left( 0\right)\equiv  \alpha ^{\prime }\left(
0\right),\\ \\
\displaystyle R_{0M}^2\left( 0\right) =R_{0B}^2\left( 0\right) =R_{0D}^2\left(
0\right)  \equiv  R_{0}^2\left( 0\right),\\ \\
\left( N_MN_D\right) ^{1/2}=N_B .
\end{array}
\end{equation}
Then the relations (\ref{OurPlanar}) and (\ref{NormCnst}) can be
fulfilled at all $s$. Otherwise, they can only be satisfied at
sufficiently large $s$ (cf. Ref. (\cite{Kaidalov1})).

\section{Deuteron photodisintegration in the QGSM}\label{spin}
Before going over to the case of particles with spin we first
present the amplitudes for spinless constituents.

\subsection{Spinless particles}
Using the same approach as in previous Section we now consider the
reaction
\begin{equation}
\gamma d\rightarrow pn.
\end{equation}
By analogy to Eq. (\ref{factorization}) the amplitude
corresponding to each quark diagram of Fig. 1 can be written as
\begin{equation}\label{factorizgamd}
\widetilde{A}^{\gamma d\to pn}(s,{\bf b}_{\perp })=\frac
i{2s}\ \widetilde{T}^{\gamma d\to q (5q)}(s,{\bf b}_{\perp })\ \widetilde{T}
^{q (5q)\to pn}(s,{\bf b}_{\perp }),
\end{equation}
where the amplitudes  $\widetilde{T}^{\gamma d\to q (5q)}(s,{\bf
b}_{\perp })$ and $\widetilde{T}^{q (5q)\to pn}(s,{\bf b}_{\perp
})$ are given by (cf. (\ref{PionNucleonFragmentation}))
\begin{equation} \label{deutFragmentation}
\begin{array}{c}
\displaystyle \widetilde{T}^{\gamma d \rightarrow q(5q)}
(s,  {\bf b}_{\perp})=N_{M(6q)}^{1/2}\frac 1{2\sqrt{\pi }R_{M(6q)}
\left( s\right)
}\left( -\frac s{m_0^2}\right) ^{\left( \alpha _M\left( 0\right) +1\right)
/2}\exp \left( -  \frac{{\bf b}_{\perp }^2}{8R_{M(6q)}^2\left( s\right) }
\right), \\
\\
\displaystyle \widetilde{T}^{q(5q)\rightarrow pn}(s,{\bf
b}_{\perp})= N_{D(6q)}^{1/2}\frac 1{2\sqrt{\pi }
R_{D(6q)}\left(s\right) }
\left( -\frac s{m_0^2}\right)^{\left( \alpha _D\left( 0\right)+1\right) /2}
\exp \left( -\frac{{\bf b}_{\perp }^2}{8R_{D(6q)}^2\left( s\right)}\right).
\end{array}
\end{equation}
Here the effective
interaction radii $R_{M(6q)}\left( s\right)$ and $R_{D(6q)}\left( s\right)$
are defined as
\begin{equation}\label{Radii1}
\begin{array}{c}
\displaystyle R_{M(6q)}^2\left( s\right)
=R_{0M(6q)}^2+\alpha _M^{\prime }\left( 0\right) \ln \left(
-\frac s{m_0^2}\right),  \\ \\
\displaystyle R_{D(6q)}^2\left( s\right) =
R_{0D(6q)}^2+\alpha _D^{\prime }\left( 0\right) \ln \left(
-\frac s{m_0^2}\right),
\end{array}
\end{equation}
where $R_{0M(6q)}^2$ and $R_{0D(6q)}^2$ are, in general, different
from $R_{0M}^2$ and $R_{0D}^2$ in Eq. (\ref{Radii}).

\subsection{Full amplitudes with spin variables}
In case of constituents with explicit spin we write the deuteron
photodisintegration amplitude in the form
\begin{eqnarray}\label{ImpReprspin}
\lefteqn{ \langle p_3, \lambda_{p}; p_4,
\lambda_{n} | \hat{T}\left(s,{\bf p}_{3 \perp}\right)| p_2,
\lambda_{d}; p_1, \lambda_{\gamma}\rangle =}  \nonumber \\
&&\frac i{8\pi ^2s}\int d^2{\bf k}_{\perp } \
\langle \lambda_{p};
\lambda_{n} | \hat{T}^{q(5q)\rightarrow pn}\left(s,{\bf k}_{\perp}\right)|
\lambda_{q};\lambda_{(5q)}\rangle\nonumber \\
&& \langle \lambda_{q}; \lambda_{(5q)} | \hat{T}^{\gamma d\rightarrow q(5q)}
\left(s,{\bf p }_{ 3 \perp}-{\bf k }_{\perp}\right)|
\lambda_{d};\lambda_{\gamma}\rangle \ ,
\label{convolutionspin}
\end{eqnarray}
where $p_1$, $p_2$, $p_3$, and $p_4$ are the 4-momenta of the
photon, deuteron, proton, and neutron, respectively, while
$\lambda_i$ is the $s$ channel helicity of the $i$-th particle.
Furthermore, we make simplifying assumption that the spin of the
$(5q)$ state is $1/2$. Then we can write the amplitude $\hat
{T}^{\gamma d\rightarrow q(5q)}$ as
\begin{eqnarray}
\lefteqn{ \langle \lambda_{q};
\lambda_{(5q)} | \hat{T}\left(s,{\bf k}_{\perp}\right)|
\lambda_{d};\lambda_{\gamma}\rangle =}  \nonumber \\
&& \bar u_{\lambda_q}(p_q) \hat {\epsilon}_{\lambda_{\gamma}}
\left(\frac{-\hat k+m_q}{k^2-m_q^2}\right)
\hat {\epsilon}_{\lambda_d} v_{\lambda_{(5q)}}(p_{(5q)}) \
D^{\gamma d\rightarrow q(5q)}(s,{\bf k}_{\perp})\ ,
\label{gamdq5qAmpl}
\end{eqnarray}
where ${\epsilon}_{\lambda_{d}}$ and
${\epsilon}_{\lambda_{\gamma}}$ are the deuteron and photon
polarization vectors, $D^{\gamma d\rightarrow q(5q)}(s,{\bf
k}_{\perp})$  is the scalar amplitude and  $m_q$ is the quark
mass. In analogy to $q\overline{q}\rightarrow N\overline{N}$,
which was analysed in Ref. \cite{Chekin}, we can describe the spin
structure of the amplitude $\hat {T}^{q(5q)\rightarrow pn}$ in
terms of eight invariant amplitudes
\begin{eqnarray}
\lefteqn{ \langle \lambda_{p};
\lambda_{n} | \hat{T}^{q+(5q)\rightarrow
pn}\left(s,{\bf k}_{\perp}\right)|
\lambda_{q};\lambda_{(5q)}\rangle =}  \nonumber \\
&&D_1(s,{\bf k}_{\perp})\ \delta_{\lambda_p \, \lambda_q}
\delta_{\lambda_n \, \lambda_{(5q)}}+
D_2(s,{\bf k}_{\perp})\ (\sigma_y)_{\lambda_p \, \lambda_q}
\delta_{\lambda_n \, \lambda_{(5q)}}+\nonumber\\
&&D_3(s,{\bf k}_{\perp})\ \delta_{\lambda_p \, \lambda_q}
(\sigma_y)_{\lambda_n \, \lambda_{(5q)}}+
D_4(s,{\bf k}_{\perp})\ (\sigma_x)_{\lambda_p \, \lambda_q}
(\sigma_x)_{\lambda_n \, \lambda_{(5q)}}+\nonumber\\
&&D_5(s,{\bf k}_{\perp})\ (\sigma_y)_{\lambda_p \, \lambda_q}
(\sigma_y)_{\lambda_n \, \lambda_{(5q)}}+
D_6(s,{\bf k}_{\perp})\ (\sigma_z)_{\lambda_p \, \lambda_q}
(\sigma_z)_{\lambda_n \, \lambda_{(5q)}}+\nonumber\\
&&D_7(s,{\bf k}_{\perp})\ (\sigma_x)_{\lambda_p \, \lambda_q}
(\sigma_z)_{\lambda_n \, \lambda_{(5q)}}+
D_8(s,{\bf k}_{\perp})\ (\sigma_z)_{\lambda_p \, \lambda_q}
(\sigma_x)_{\lambda_n \, \lambda_{(5q)}} \ ,
\label{q5qpnAmpl}
\end{eqnarray}
where the $z$- and $x$-axes are directed along the photon
momentum and the momentum transfer ${\bf k}_{\perp}$,
respectively, and the $y$-axis is orthogonal to the scattering
plane.

Now the experimental data on the proton form factor are in
agreement with the assumption, that the dominant contribution
stems from the amplitude corresponding to the conservation of the
$s$-channel helicities (cf. Ref.  \cite{Chekin}). Here we shall
use the same assumption and take into account only the amplitude
$D_1(s,{\bf k}_{\perp})$. We thus find
\begin{eqnarray}
\lefteqn{ \langle \lambda_{p};
\lambda_{n} | \hat{T}\left(s,{\bf p}_{3 \perp}\right)|
\lambda_{d};\lambda_{\gamma}\rangle =}  \nonumber \\
&&\frac i{8\pi ^2s}\int \! d^2{\bf k}_{\perp } \
 \bar u_{\lambda_p}(p_3) \hat {\epsilon}_{ \lambda_{\gamma}}
\left(\frac{-\hat k+m_q}{k^2-m_q^2}\right)
\hat {\epsilon}_{\lambda_{d}} v_{\lambda_n}(p_4) \nonumber \\
&& \times D^{\gamma d\rightarrow q(5q)}(s,{\bf k}_{\perp})
\ D_{1}(s,{\bf p}_{3 \perp}-{\bf k}_{\perp}).
\label{convolutionspin1}
\end{eqnarray}
Furthermore, taking into account that at high energy
$p_{\gamma}\gg \sqrt{s_0}$ and finite momentum transfer $t\simeq
|{\bf p}_{3\,\perp}|^2\simeq s_0$ the momentum $k$ is almost
transversal $k=(k_0,{\bf k}_{\perp},k_z)$, where $\displaystyle
k_0\simeq k_z\simeq O\left(\frac{s_0}{2 p_{\gamma}}\right)$ and
$\int \! d^2{\bf k}_{\perp}\ {\bf k }_{\perp}(...) \sim {\bf
p}_{3\,\perp}$, we find the following representation for the spin
structure of the $\gamma d\rightarrow pn$ amplitude:
\begin{eqnarray}
\lefteqn{ \langle \lambda_{p};
\lambda_{n} | \hat{T}\left(s,{\bf p}_{3 \perp}\right)|
\lambda_{d};\lambda_{\gamma}\rangle =}  \nonumber \\
&& \bar u_{\lambda_p}(p_3) \hat {\epsilon}_{\lambda_{\gamma}}
\left({A(s,t){\bf p}_{3\,\perp}\cdot \mbox{\boldmath $\gamma$}
+B(s,t) m}\right)
\hat {\epsilon}_{\lambda_{d}} v_{\lambda_n}(p_4)\ ,
\label{spin1}
\end{eqnarray}
where
\begin{eqnarray}
&&A(s,t)=\frac i{8\pi ^2s}\int \! d^2{\bf k}_{\perp } \
\frac{{\bf k}_{\perp}\cdot{\bf p}_{3\,\perp}}
{|{\bf p}_{3\,\perp}|^2}\
\frac{1}{k^2-m_q^2}  \nonumber \\
&&\times D^{\gamma d\rightarrow q(5q)}(s,{\bf k}_{\perp})
\ D_{1}^{q(5q)\rightarrow pn}(s,{\bf p}_{3\, \perp}-{\bf k}_{\perp})\ ,
\label{convolutionscalarA}\\
&&B(s,t)=\frac i{8\pi ^2s} \frac{m_q}{m}\int \! d^2{\bf k}_{\perp } \
\frac{1}{k^2-m_q^2}  \nonumber \\
&&\times D^{\gamma d\rightarrow q(5q)}(s,{\bf k}_{\perp})
\ D_{1}^{q(5q)\rightarrow pn}(s,{\bf p}_{3\, \perp}-{\bf k}_{\perp})\  ,
\label{convolutionscalarB}
\end{eqnarray}
and $m$ is the nucleon mass. In the case of a Gaussian parametrization
for  $ D^{\gamma d\rightarrow q(5q)}(s,{\bf k}_{\perp})$ and $D_1$
(in Eqs. (24,25)) the ratio $R={A(s,t)}/{B(s,t)}$
is a smooth function of $t$. We, furtheron, assume that it is a
constant and consider this constant as a free parameter.

The differential cross section for the reaction $\gamma
d\rightarrow pn$ then is
\begin{eqnarray}
&\displaystyle \frac{d\sigma^{I}_{\gamma d\to pn}}{d t}& =
\frac{1}{64\,\pi s}\ \frac{1}{(p_{\gamma}^{\mathrm{cm}})^2}\
\left[S_{t}\ |B(s,t)|^2+S_u\ |B(s,u)|^2 \right. \nonumber\\
&&\left. +(-1)^{I+1}\  2S_{tu}\
{\mathrm Re}(B(s,t)B(s,u)\right],
\label{eq:sigt}
\end{eqnarray}
where~$I$ is the isospin of the reaction, i.e. $I=1$ (or $0$) for
isovector (or isoscalar) photons. The kinematical functions $S_t$,
$S_u$, $S_{tu}$ in (\ref{eq:sigt}) are given by
\begin{eqnarray}
S_t=\frac{1}{6}\  \sum _{\lambda_{\gamma},\ \lambda_{d}}
{\mathrm {Sp}}\ \left[ \hat {\epsilon}_{\lambda_{\gamma}}
\left(R \,{\bf p}_{3\,\perp}\cdot \mbox{\boldmath $\gamma$}
+ m \right)
\hat {\epsilon}_{\lambda_{d}}
\left(\hat {p}_4- m  \right) \right.  \nonumber \\
\times \left. \hat {\epsilon}^{\star}_{\lambda_{d}}
\left(R \, {\bf p}_{3\,\perp}\cdot \mbox{\boldmath $\gamma$}
+  m \right)
 \hat {\epsilon}^{\star}_{\lambda_{\gamma}}
\left(\hat {p}_3+ m  \right) \right] \ ,\nonumber \\
S_u=\frac{1}{6}\  \sum _{\lambda_{\gamma},\ \lambda_{d}}
{\mathrm {Sp}}\  \left[ \hat {\epsilon}_{\lambda_{d}}
\left(R\, {\bf p}_{3\,\perp}\cdot \mbox{\boldmath $\gamma$}
+  m \right)
 \hat {\epsilon}_{\lambda_{\gamma}}
\left(\hat {p}_4- m  \right) \right. \nonumber \\ \times \left.
\hat {\epsilon}^{\star}_{\lambda_{\gamma}} \left(R\, {\bf
p}_{3\,\perp}\cdot \mbox{\boldmath $\gamma$} +  m \right) \hat
{\epsilon}^{\star}_{\lambda_{d}} \left(\hat {p}_3+ m
\right)\right] \ ,\nonumber\\ S_{tu}=\frac{1}{6}\  \sum
_{\lambda_{\gamma},\ \lambda_{d}} {\mathrm {Sp}}\ \left[ \hat
{\epsilon}_{\lambda_{\gamma}} \left(R \, {\bf p}_{3\,\perp}\cdot
\mbox{\boldmath $\gamma$} + m \right) \hat
{\epsilon}_{\lambda_{d}} \left(\hat {p}_4- m  \right)\right.
\nonumber \\ \times \left. \hat
{\epsilon}^{\star}_{\lambda_{\gamma}} \left(R\, {\bf
p}_{3\,\perp}\cdot \mbox{\boldmath $\gamma$} + m \right) \hat
{\epsilon}^{\star}_{\lambda_{d}} \left(\hat {p}_3+ m
\right)\right] \ .
\end{eqnarray}
In order to fix the energy dependence of the amplitude $B(s,t)$ we
require that \begin{equation}  \displaystyle \left.\frac{{\mathrm
d}\sigma}{{\mathrm d }t} \right|_{\; \Theta_{{\mathrm
c.m.}}=0}\sim \left(\frac{s}{s_0}\right)^{2\alpha_N(0)-2} \ .
\end{equation} Taking into account that $S_{t}\sim s$ for $s\gg s_0$ we
find that \begin{equation}
 \displaystyle
B(s,t)\sim\left(\frac{s}{s_0}\right)^ {\alpha_N(0)-1/2} \ .
\end{equation} Moreover, a good approximation for the energy dependence
of $S_t(\theta_{{\mathrm c.m.}}=0)$ in the region $p_{\gamma}=
1\div 7.5$~GeV is \begin{equation} \left.S_t \right|_{\;
\theta_{{\mathrm c.m.}}=0}\approx C p_{\gamma}^2 \end{equation}
with $C=(36\pm 3)~$GeV$^2$. Using this approximation we can relate
$B(s,t)$ to the Regge-pole exchange amplitude as
\begin{equation}
|B(s,t)|^{2}=\frac{1}{C p_{\gamma}^2}\
|\mathcal{M}_{{\mathrm Regge }}(s,t)|^2\ ,
\label{Bst}
\end{equation}
where
\begin{equation}
\mathcal{M}_{{\mathrm Regge}}(s,t)= F(t)
\left(\frac{s}{s_0}\right)^{\alpha_{N}(t)} \exp{\left[
      -i\ \frac{\pi}{2}\left(\alpha_{N}(t) -
        \frac{1}{2}\right)\right]}\  .
\label{eq:Mregge}
\end{equation}
Here $\alpha_N(t)$ is the trajectory of the nucleon Regge pole and
$s_0 =4~\mathrm{GeV}^2 \simeq m_d^2$.

\subsection{Nonlinear nucleon Regge trajectories}
According to the data on $\pi N$ backward scattering (see e.g. the
review \cite{Lyubimov}) the nucleon Regge trajectory has a
nonlinearity:
\begin{equation}
  \alpha_{N}(t)= \alpha_{N}(0) +\alpha'_{N}(0)\, t +
  \frac{1}{2}\, \alpha''_{N}(0)\, t^2 \, K(t),
\label{eq:alpha1}
\end{equation}
where $\alpha_{N}(0)=-0.5$, $\alpha'_{N}(0)=0.9\
\mathrm{GeV}^{-2}$ are the intercept and slope of the Regge
trajectory, $\alpha''_{N}(0)= 0.20 \div 0.25 \mathrm{GeV}^{-4}$ is
the coefficient of the nonlinear term. In (\ref{eq:alpha1}) we
introduced also a cut-off function $K(t)$. In Ref. \cite{Lyubimov}
it was assumed that $K(t) =1$. However, in this case the amplitude
will grow very fast with $s$ at large $t$ which would violate
unitarity. To prevent this fast growth it was taken in Ref.
\cite{Desanctis} as a powerlike cut-off $ K(t)=(1+ {t^4}/{{\Lambda
}^4})^{-1}$. Here we choose the exponential form
\begin{equation}
K(t)=\exp\left({-\beta t^2}\right) \end{equation}
with $\beta =
0.008\, \mathrm{GeV}^{-4}$. We mention that the small value of
$\beta$ does not destroy the parameterization of $\alpha(t)$ for
$-t \leq 1.6\ \mathrm{GeV}^2$ found in Ref. \cite{Lyubimov}. Note
also that the phenomenological Regge trajectory (\ref{eq:alpha1})
with a powerlike or exponential cut-off is nonlinear only for
moderate values of $t$; at large $t$ the quadratic term becomes
small and the trajectory becomes essentially linear again.

On the other hand, the QCD motivated Regge trajectories as
suggested by Brisudov\'{a}, Burakovsky and Goldman (BBG)
\cite{Burak} show a different behaviour at large $t$. As  shown in
Ref. \cite{Burak} the screened quark-antiquark potential
\begin{equation}
V(R)=\left[ -\frac{\alpha}{R}+\sigma R \right] \frac{1-\exp (- \mu R)}{\mu R}
\end{equation}
with $\sigma=$(400~MeV)$^2$, $\mu^{-1} =0.90 \pm 0.20$~fm, $\alpha
=0.21\pm0.1$, found in Ref. \cite{Born} to describe the lattice
QCD data with dynamical Kogut-Susskind fermions, leads to
nonlinear meson Regge trajectories. These trajectories can be
parametrized on the whole physical sheet as
\begin{equation}
  \alpha(t) = \alpha(0)+ \gamma \left[ T^{\nu} - \left(T-t\right)^{\nu}
\right] \label{eq:nonlinBBG}
\end{equation}
with $0\leq \nu \leq 1/2$. The limiting cases $\nu=1/2$ and $\nu
\to 0$ ($\gamma \nu =$ const) correspond to the square-root
trajectory
\begin{equation}
  \alpha(t) = \alpha(0)+ \gamma \left[ \sqrt{T} - \sqrt{T-t}
\right],
\label{eq:sqroot}
\end{equation}
and the logarithmic trajectory
\begin{equation}
  \alpha(t) = \alpha(0)- (\gamma \nu) \ln\left(1 - \frac{t}{T}
\right) \label{eq:log}
\end{equation}
respectively. Such trajectories arise not only for heavy
quarkonia, but also for light-flavour hadrons.

To find the possible form of nonlinear  Regge trajectories for
mesons composed of light quarks,  Brisudov\'{a}, Burakovsky and
Goldman \cite{Burak} have considered an analytical model for a
string with massless ends and variable string tension. This model
describes a colour flux tube stretched between quark and antiquark
at the tube ends. The varying string tension was introduced to
simulate dynamical effects such as the weakening of the flux tube
due to pair ($q \bar{q}$) creation. Within this model they were
able to recover the form of the underlying potential for a given
 Regge trajectory. They found potentials leading to
``square-root'' and ``logarithmic'' Regge trajectories and
demonstrated, that these potentials are very similar to the
screened potential of the unquenched lattice QCD calculations.
Moreover, they were able to describe very well all the available
meson spectra using a square-root Regge trajectory. Nevertheless,
the ``logarithmic'' form of Regge trajectories can not be excluded
by now and new data on higher excited states are necessary.

We know from experiment that the slopes of meson and baryon Regge
trajectories are almost the same $\alpha^{\prime}_N \simeq
\alpha^{\prime}_{\rho} \simeq 0.9-1 $ GeV$^{-2}$ (see e.g. the
review \cite{KaidalovSurvey}). The slope is determimed by the string
tension which depends on the colour charges at the string ends.
Therefore, the baryon Regge trajectory can be described by the
colour flux model with quark and diquark at the ends (cf Ref.
\cite{Kobzarev}). This means that for the baryon Regge
trajectories we can also use the forms suggested by BBG.

When using the QCD motivated trajectories (\ref{eq:sqroot}) and
(\ref{eq:log}) we take their intercepts and slopes the same as in
case of the phenomenological trajectories. Therefore, only a
single free parameter will be left. We choose $T=T_B$ as this free
parameter and fix it by comparing our results to experimental
data.

\section{Cross sections for deuteron photodisintegration in the QGSM}
\label{sec:results}

\subsection{Choice of parameters}

The dependence of the residue $F(t)$ on $t$ can be taken from
Refs. \cite {Kaidalov4,Guaraldo} in the form
\begin{equation}
  F(t) = B {\left[\frac{1}{m^2 - t}\ \exp{(R_1^2t)} + C\, \exp{(R_2^2
        t)} \right]}\ ,
\label{eq:resid1}
\end{equation}
where the first term in the square brackets contains the nucleon
pole and the second term accounts for the contribution of
non-nucleonic degrees of freedom in the deuteron.  We adopted the
following sets of parameters:\\ i) for the case of the
phenomenological nonlinear trajectory
\begin{eqnarray}
&&\mathrm{Set\ K \!:}\ \;
B=3.32\cdot10^{-4}\, \mathrm{kb}^{1/2}\cdot\mathrm{GeV},\;
C = 0.7\ \mathrm{GeV}^{-2}  , \nonumber\\
&&\hphantom{Set\ K \!:}\ R_1^2 = 1\ \mathrm{GeV}^{-2} ,  \;
R_2^2 =- 0.1\  \mathrm{GeV}^{-2} , \;
\alpha''_{N(0)}= 0.20\, \mathrm{GeV}^{-4}
\: , \nonumber \\
&&\mathrm{Set\ G \!:}\ \;
B=4.01\cdot10^{-4}\, \mathrm{kb}^{1/2}\cdot\mathrm{GeV} ,\;
C = 0.7\ \mathrm{GeV}^{-2}  , \nonumber\\
&&\hphantom{Set\ K \!:}\ R_1^2 = 2\ \mathrm{GeV}^{-2} , \;
R_2^2 = 0.03\  \mathrm{GeV}^{-2}
 ,\;
\alpha''_{N(0)}= 0.25\, \mathrm{GeV}^{-4}
\:
\end{eqnarray}
and the ratio $R=A(s,t)/B(s,t)=1$ for both sets K and G. The
parameters of the residue (\ref{eq:resid1}) in $\mathrm{Set\ K}$,
except the overall normalization factor $B$ and $R^2_1$, are the
same as in Ref. \cite{Kaidalov4} which were fitted to data on the
reaction $pp \to \pi^+ d$ for $-t \leq 1.6\ \mathrm{GeV}^2$.
Therefore, in this case we have only two free parameters of the
residue $B$ and $R_1^2$ which we fixed by the experimental data on
deuteron photodisintegration at
$\Theta_{\mathrm{c.m.}}=36^{\circ}$. We note that $\mathrm{Set\
G}$ corresponds to positive values of $R_2^{2}$;\\ ii) for the
case of the QCD motivated Regge trajectories we have used the
parameters of the residue from Set~G except an overall
normalization factor  $B$ taken as $B=1.8 \cdot 10^{-4}$ kb$^{1/2}
\cdot $ GeV for the logarithmic and $B=2.0 \cdot 10^{-4}$
kb$^{1/2} \cdot $ GeV for the square-root trajectory. Furthermore,
we choose $R=A(s,t)/B(s,t)=2$ for both trajectories (logarithmic
and square-root).  In order to achieve a good agreement with the
data at $\Theta_{\mathrm{c.m.}}=36^{\circ}$ the nonlinearity
parameter $T=T_B$ was chosen in the interval 1.5-1.7 GeV$^2$.

\subsection{Energy dependence of the differential cross section}

\begin{figure}[gd36:89]
  \begin{center}
    \leavevmode
    \psfig{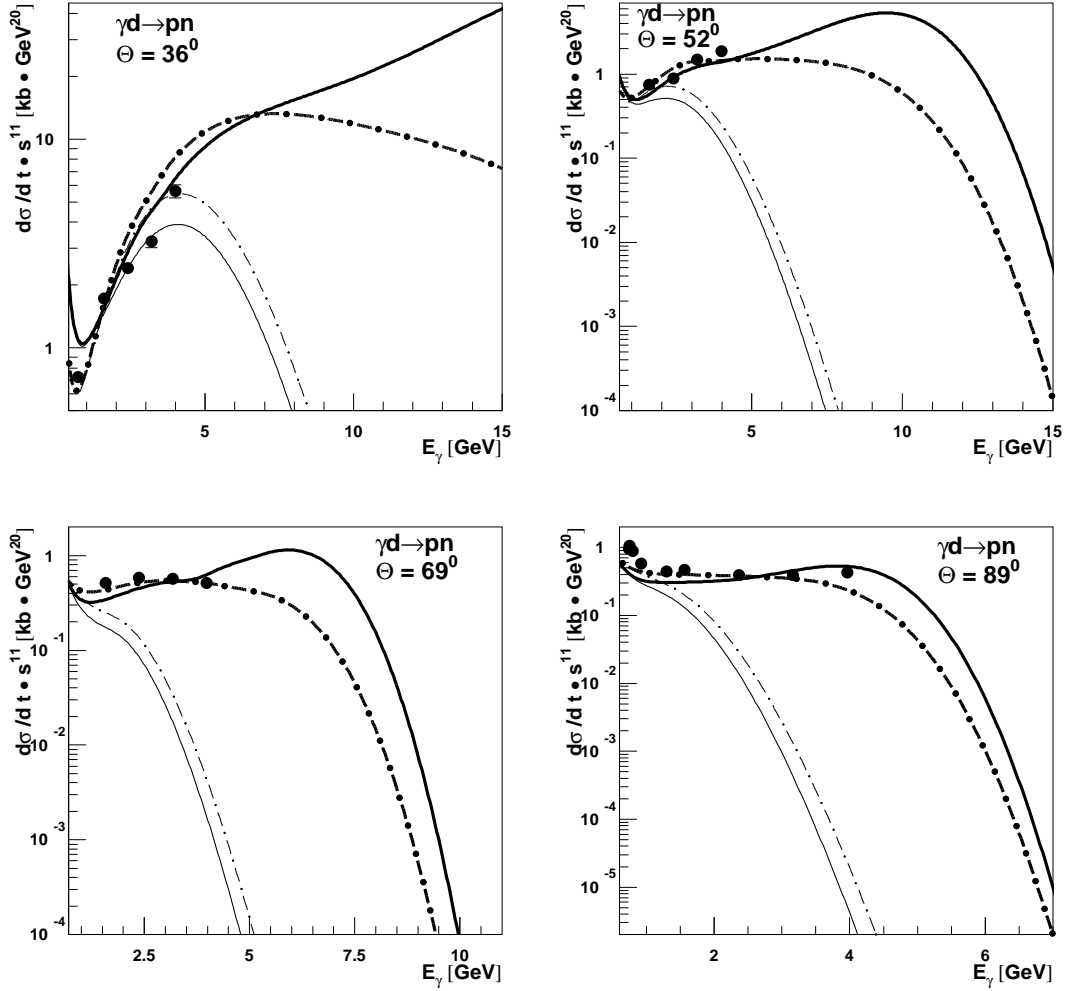}
    \caption{Differential cross section for the
      reaction $\gamma d\to pn$ (multiplied
by $s^{11}$) as a function of the photon lab. energy $E_{\gamma}$
at different angles in the center-of-mass frame in comparison to
the experimental data from Ref. \cite{Holt}. The bold solid and
dash-dotted curves present results of calculations using a
phenomenological Regge trajectory for parameter Set G and Set K,
respectively. The thin lines show the results obtained in the case
of the linear Regge trajectories, i.e. for $\alpha''_{N(0)}= 0$,
with the same parameters of the residue as in Set G (solid line)
and Set K (dash-dotted line).} \label{fig:gd36:89}
  \end{center}
\end{figure}

In Fig. \ref{fig:gd36:89} we present the energy dependence of
${d\sigma}/{dt}\cdot s^{11}$ at different c.m. angles calculated
for the case of the phenomenological nonlinear trajectory
(\ref{eq:alpha1}). The experimental points are taken from
\cite{Holt}. The bold solid and dash-dotted lines present results
of calculations within the QGSM for parameters of the Set G and
Set~K, respectively. The thin lines show the results obtained in
the case of the linear Regge trajectories with residue parameters
from Set~G (thin solid line) and Set  K (thin dash-dotted line).
For $\Theta_{\mathrm{c.m.}}=36^{\circ}$ (see the top-left part)
all the curves except the thin solid line are in a reasonable
agreement with the data. Therefore, for this angle it is also
possible to describe the data using a linear nucleon Regge
trajectory. However, at large angles
$\Theta_{\mathrm{c.m.}}=52^{\circ}$, $69^{\circ}$ and $89^{\circ}$
the nonlinearity becomes essential since the thin curves
underestimate the experimental points substantially. The use of
the nonlinear Regge trajectory instead provides a reasonable
description of the existing data and reproduces the scaling
behavior of $d\sigma/dt\cdot s^{11}$ for
$\Theta_{\mathrm{c.m.}}=69^{\circ}$ and $89^{\circ}$ at energies
$E_{\gamma}\le 5$~GeV. At higher energies all curves drop very
fast.

\begin{figure}[gd36:89b]
  \begin{center}
    \leavevmode
    \psfig{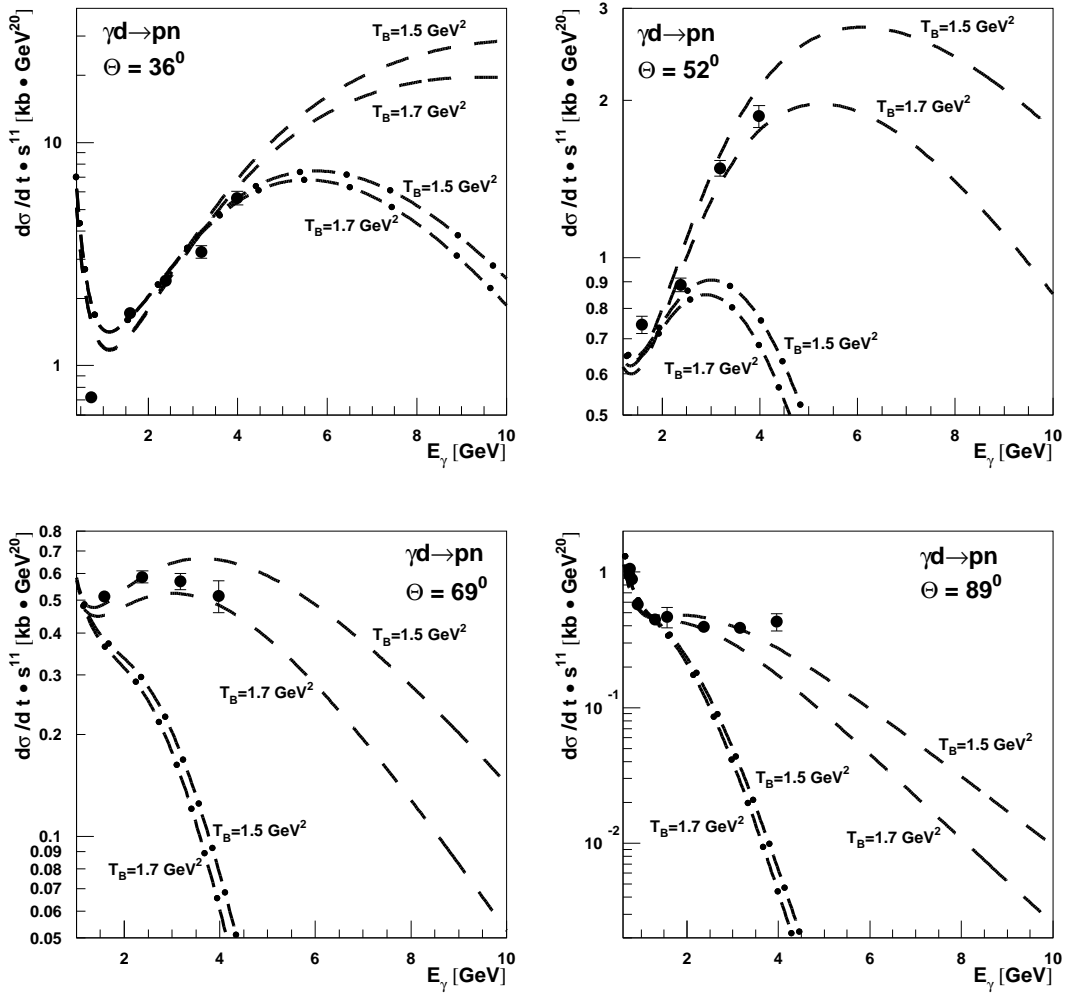}
    \caption{Differential cross section for the
      reaction $\gamma d\to pn$ (multiplied
by $s^{11}$) as a function of the photon lab. energy $E_{\gamma}$
at different angles in the center-of-mass frame in comparison to
the experimental data from Ref. \cite{Holt}. The dashed and
dash-dotted curves are calculated using the logarithmic and
square-root Regge trajectories, respectively.}
 \label{fig:gd36:89b}
 \end{center}
\end{figure}

In Fig. \ref{fig:gd36:89b} the energy dependence of
${d\sigma}/{dt}\cdot s^{11}$ at different c.m. angles is
calculated for the cases of the square-root (\ref{eq:sqroot})
(dash-dotted lines)
 and logarithmic (\ref{eq:log}) (dashed lines) Regge trajectories.
The lower and upper curves were calculated with $T_B=$1.7 and 1.5
GeV$^2$, respectively. As noted before, the ratio of the invariant
amplitudes $R=A(s,t)/B(s,t)$ is taken as $R$= 2. It is seen that
the result of the square-root trajectory leads to a cross section
which underestimates the data for $\Theta_{\mathrm{c.m.}}\geq
52^{\circ}$, while the logarithmic trajectory provides a
reasonable description of the data at all angles (with the
exception of a single point at
$\Theta_{\mathrm{c.m.}}=89^{\circ}$, $E_{\gamma}= 4$ GeV).
Therefore, new measurements of $d\sigma/dt$ at $E_{\gamma}\ge
5$~GeV will provide a crucial check of the QGSM predictions.

\subsection{Angular dependence of the cross section}

\begin{figure}[dsg]
  \begin{center}
    \leavevmode
    \psfig{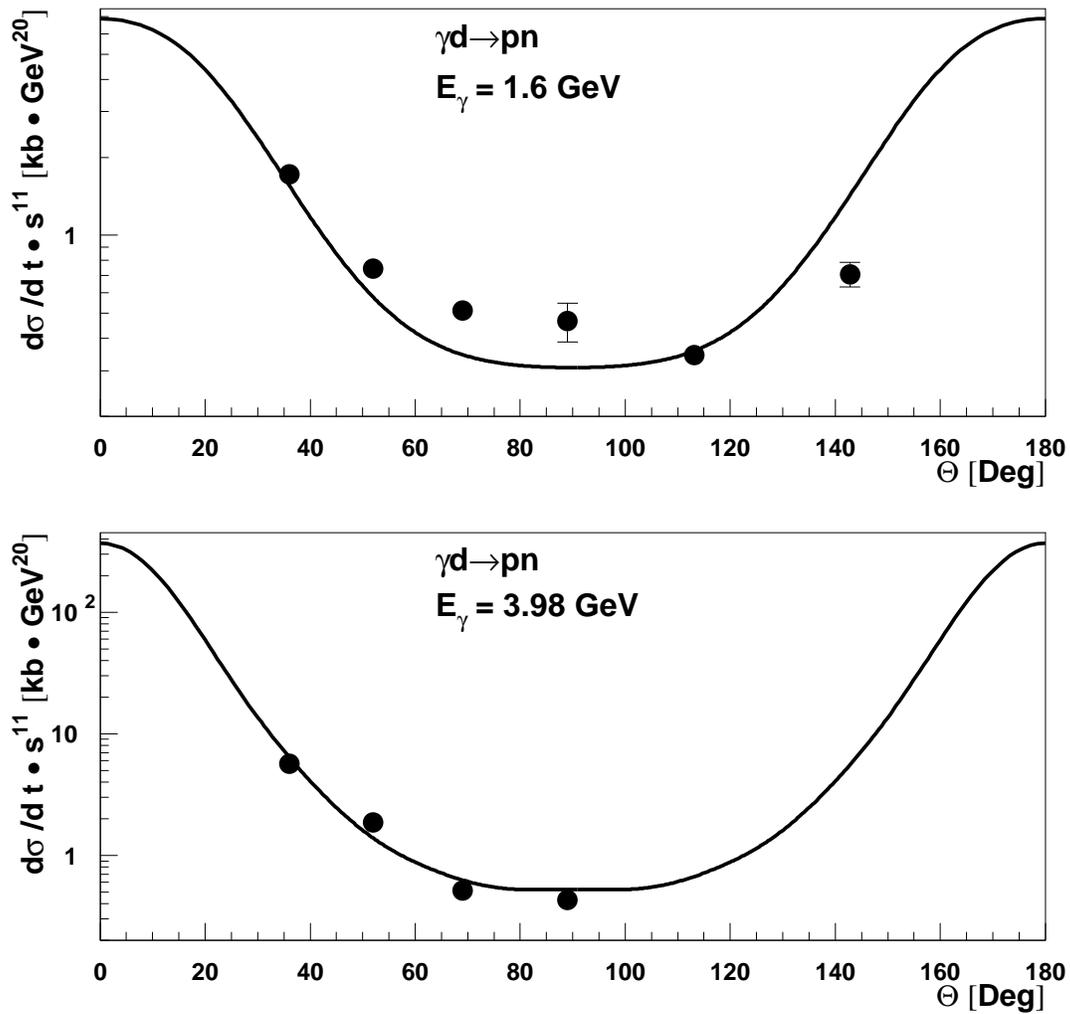}
    \caption{Differential cross section for the
      reaction $\gamma d\to pn$ (multiplied
by $s^{11}$) as a function of the c.~m. angle for
$E_{\gamma}=1.6$~GeV and 3.98~GeV (upper and lower parts,
respectively). The experimental data are from Ref. \cite{Holt}.
The bold solid curves are results of calculations using the
phenomenological Regge trajectory  for parameter Set G.}
\label{fig:dsg}
  \end{center}
\end{figure}

In Fig. \ref{fig:dsg} we show  the angular  dependence of the
differential cross section as a function of
$\Theta_{\mathrm{c.m}.}$ calculated for $E_{\gamma}$ =1.6 GeV and
3.98 GeV using the phenomenological nonlinear trajectory
(\ref{eq:alpha1}) with  $\alpha ^{\prime \prime}(0) = 0.25 $
GeV$^{-4}$ and residue parameters of the set G. Here we have
assumed isovector photon dominance and therefore obtain a
forward-backward symmetric cross sections. Both angular
distributions have forward and backward peaks, which are mainly
related to the choice $R=A(s,t)/B(s,t)$=1 in this case. The
agreement between data and calculations is fairly good.

\subsection{Forward-backward asymmetry}
\begin{figure}[dsgb1]
  \begin{center}
    \leavevmode
    \psfig{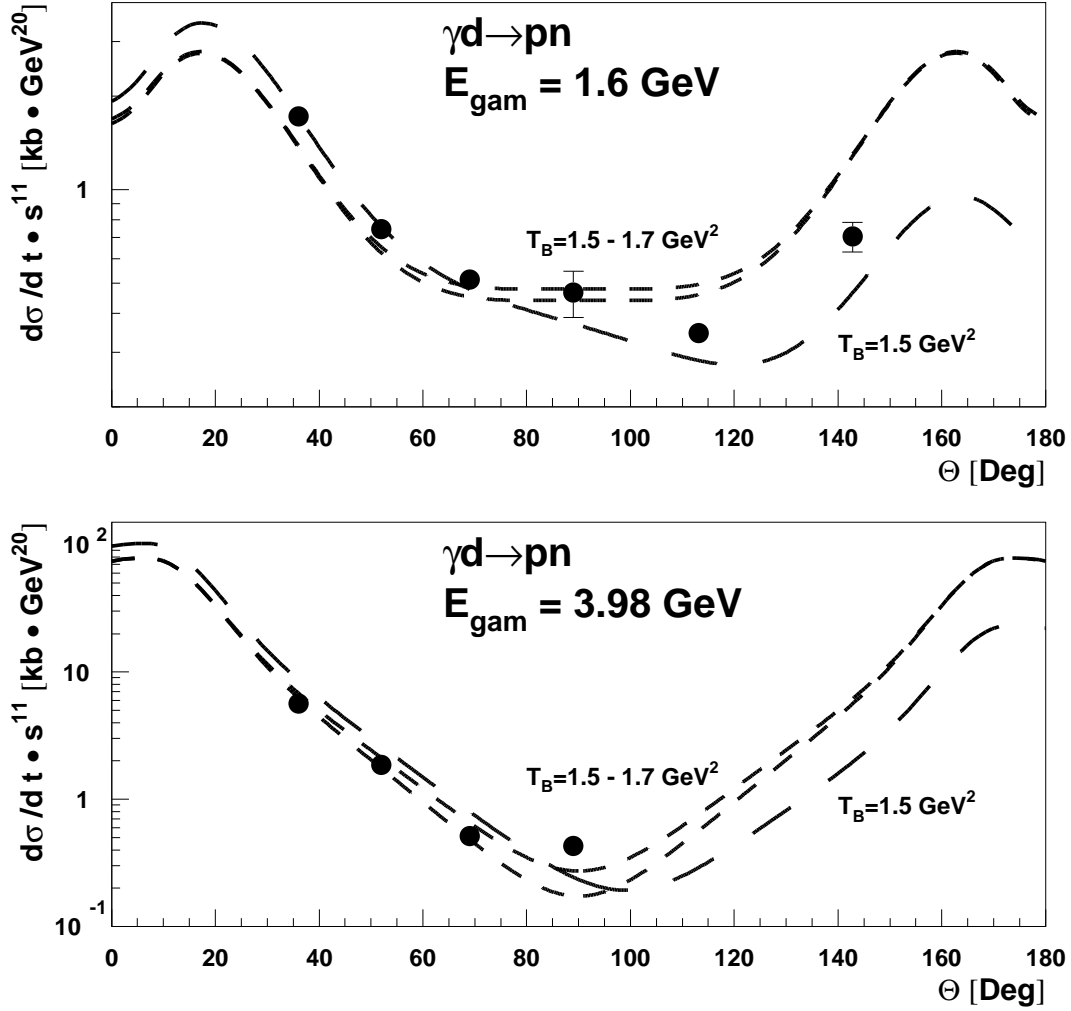}
    \caption{Differential cross section for the
      reaction $\gamma d\to pn$ (multiplied
by $s^{11}$) as a function of the c.~m. angle for
$E_{\gamma}=1.6$~GeV and 3.98~GeV (upper and lower parts,
respectively). The experimental data are from Ref. \cite{Holt}.
The dashed curves are calculated  using logarithmic Regge
trajectories with $T_B = 1.5$ and 1.7 GeV$^2$.
 The long-dashed curve presents the result
of calculations which take into account the interference of the
isoscalar and isovector parts of the $\gamma d\to pn$ amplitude
(see text).} \label{fig:dsgb1}
  \end{center}
\end{figure}

In Fig. \ref{fig:dsgb1} we present the angular  dependence of
${d\sigma}/{dt}\cdot s^{11}$ at different energies for the
logarithmic (\ref{eq:log})    Regge trajectory. The lower and
upper parts correspond to $E_{\gamma}$ = 3.98 and 1.6  GeV,
respectively. The two dashed curves in each figure were calculated
assuming isovector photon dominance. In this case we have a
forward-backward symmetry of the differential cross section. At
1.6 GeV the angular distribution has a dip at
$\Theta_{\mathrm{c.m.}}$ =0$^{\circ}$ and 180$^{\circ}$ which is
related to the different choice of the ratio $R~ (R=2)$ as
compared to the previous case $(R=1)$.

A forward-backward asymmetry arises when we take into account the
interference of two amplitudes which describe the contribution of
isovector ($\rho$ like) and isoscalar ($\omega $ like) photons. In
this case the differential cross section can be written as
\begin{eqnarray}
&\displaystyle \frac{d\sigma^{\rho+\omega}_{\gamma d\to pn}}{d t}& =
\frac{1}{64\,\pi s}\ \frac{1}{(p_{\gamma}^{\mathrm{cm}})^2}\
\left[S_{t}\ |B^{\rho}(s,t)+B^{\omega}(s,t)|^2+ \right. \nonumber\\
&&S_u\
|B^{\rho}(s,u)-B^{\omega}(s,u)|^2  \nonumber\\
&&\left. +  2S_{tu}\
{\mathrm Re}\left(B^{\rho}(s,t)+B^{\omega}(s,t)\right)^{\star}
\left(B^{\rho}(s,u)-B^{\omega}(s,u)\right)\right].
\label{eq:sigtas}
\end{eqnarray}
Using the vector dominance model we take
\begin{equation}
B^{\omega}(s,t)=B^{\rho}(s,t)/\sqrt{8}, \hspace{1cm}
B^{\omega}(s,u)=B^{\rho}(s,u)/\sqrt{8}. \end{equation} The data at
1.6 GeV provide an evidence for a forward-backward asymmetry
because the values of the differential cross sections at backward
angles are  smaller than  for the corresponding angles in the
forward region. The predictions of the simple VDM model with $\rho
-\omega$  interference are in qualitative agreement with the data
(long-dashed curves). However, for definite conclusions we will
need more systematic data at smaller angles in the forward and
also in the backward regions.

\section{Conclusions}
\label{sec:conc}

In this work we have analyzed deuteron photodisintegration at GeV
energies within the framework of the Quark-Gluon Strings Model
with nonlinear Regge trajectories. We have taken into account spin
effects assuming the dominance of those amplitudes that conserve
$s$-channel helicities. We have found that the QGSM provides a
reasonable description of the new TJNAF data on deuteron
photodisintegration at large momentum transfer $t$ and that the
energy dependence of ${\mathrm{d}} {\sigma}/{\mathrm d}t$ at
$\Theta_{\mathrm{c.m.}} = 36\div 90^{\circ}$ provides new evidence
for a nonlinearity of the Regge trajectory $\alpha_N(t)$. The best
agreement with the data can be achieved using the QCD motivated
logarithmic form of the Regge trajectory. However, new data at
larger energies will be important to further check
the energy behaviour of
$d\sigma/dt $ at fixed c.m. angles as predicted by the QGSM.

We have also investigated the angular dependence of the cross
section at different energies. The differential cross section may
have a dip at forward angles if the amplitude with a charge-like
photon coupling ($A(s,t)$ ) is dominant. By introducing the
interference of isovector and isoscalar photon contributions we
have calculated the forward-backward asymmetry of the cross
section, which can be quite pronounced at $E_{\gamma}$ = 1.6 GeV,
but will be a decreasing function of energy. New data for small
and large angles (forward and backward) will be important to check these
predictions.

\pagebreak

\pagebreak

\pagebreak

\pagebreak

\pagebreak

\pagebreak
\end{document}